\documentclass[final,12pt,authoryear]{elsarticle}

\usepackage{amssymb}
\usepackage{amsthm}

\usepackage{longtable}
\usepackage{afterpage}
\usepackage{amsmath}
\usepackage{mathtools}
\usepackage{bm}
\usepackage{xcolor}
\usepackage{url}

\journal{Computers \& Chemical Engineering}

\begin{document}

\begin{frontmatter}



\title{Convex Envelope Method for determining liquid multi-phase equilibria in systems with arbitrary number of components}

\author[add1]{Quirin Göttl\corref{cor1}}
\author[add2,add3]{Jonathan Pirnay}
\author[add2,add3,add4]{Dominik G. Grimm}
\author[add1]{Jakob Burger}

\address[add1]{Technical University of Munich, Campus Straubing for Biotechnology and Sustainability, Laboratory of Chemical Process Engineering, Schulgasse 16, 94315 Straubing, Germany.}

\address[add2]{Technical University of Munich, Campus Straubing for Biotechnology and Sustainability, Bioinformatics, Schulgasse 22, 94315 Straubing, Germany.}

\address[add3]{Weihenstephan-Triesdorf University of Applied Sciences, Petersgasse 18, 94315 Straubing, Germany.}

\address[add4]{Technical University of Munich, Department of Informatics, Boltzmannstraße 3, 85748 Garching, Germany.}

\cortext[cor1]{Corresponding author e-mail address: quirin.goettl@tum.de}



\begin{abstract}
The determination of liquid phase equilibria plays an important role in chemical process simulation. This work presents a generalization of an approach called the convex envelope method (CEM), which constructs all liquid phase equilibria over the whole composition space for a given system with an arbitrary number of components. For this matter, the composition space is discretized and the convex envelope of the Gibbs energy graph is computed. Employing the tangent plane criterion, all liquid phase equilibria can be determined in a robust way. The generalized CEM is described within a mathematical framework and it is shown to work numerically with various examples of up to six components from the literature.
\end{abstract}

\begin{graphicalabstract}
\includegraphics[width=0.9\linewidth]{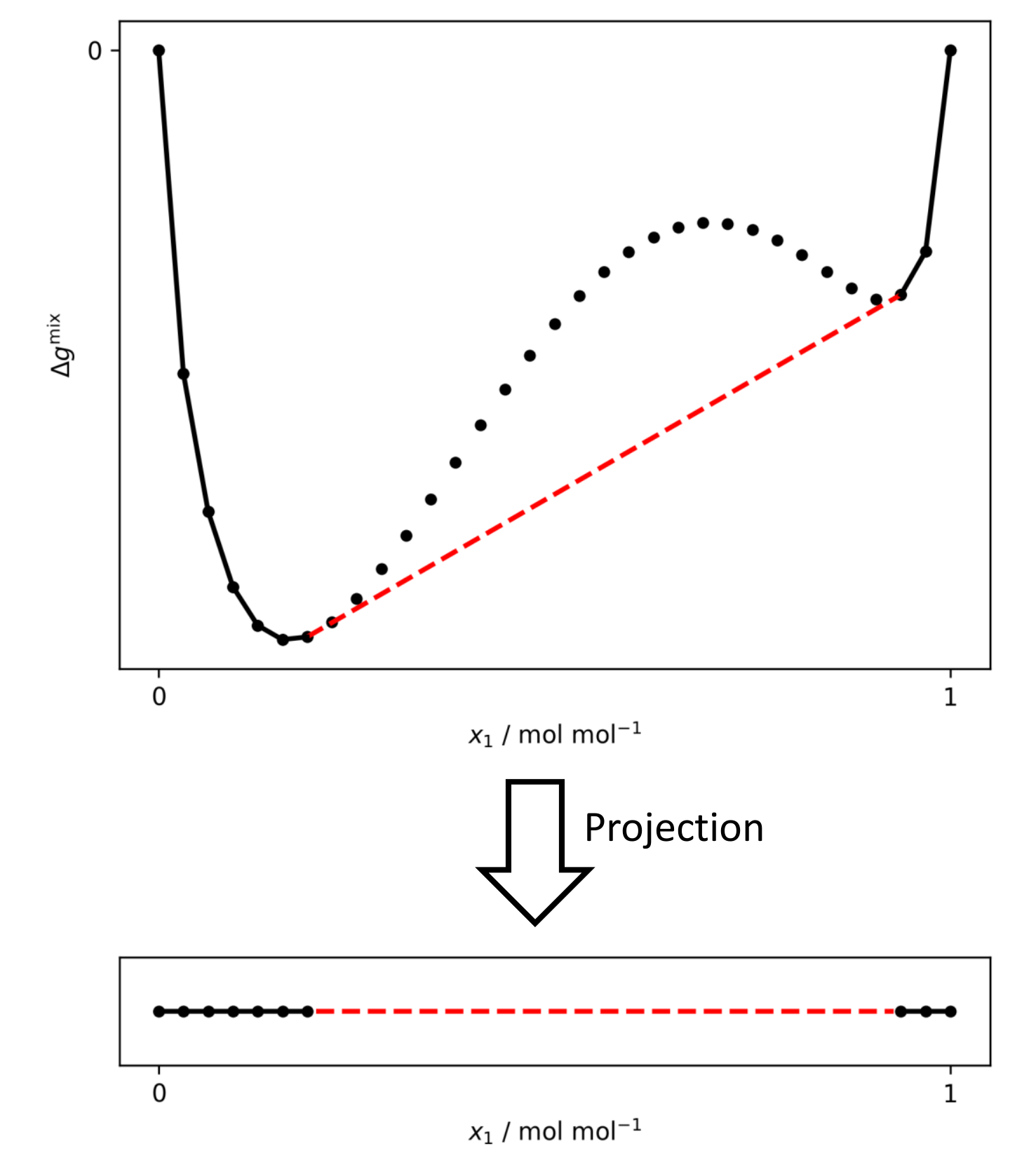}
\end{graphicalabstract}

\begin{highlights}
\item Theoretical framework for generalization of convex envelope method including a constructive proof, which enables the phase split calculations.
\item Numerical results for liquid phase equilibrium predictions for systems containing up to six components.
\end{highlights}


\end{frontmatter}


\theoremstyle{definition}
\newtheorem{defn}{Definition}
\newtheorem{theo}{Theorem}
\newtheorem{rema}{Remark}
\newtheorem{lemm}{Lemma}

\section{Introduction}
\label{introduction}
Modelling of multiphase liquid equilibria plays an important role in various research areas as chemical engineering \citep{zhang2011} or biology \citep{hyman2014}. Possible applications include design and optimization of processes for the production of biofuels \citep{merzougui2015, reynel-avila2019} or azeotropic distillation \citep{skiborowski2015}. The increase of computational power enabled machine learning methods to be used for the prediction of liquid phase equilibria \citep{moghadam2016, reynel-avila2019}. For example, surrogate models \citep{mcbride2019}, which model the underlying thermodynamics as black box models \citep{fahmi2012, nentwich2019}, are used to replace classical approaches. However, most approaches are still based on mechanistic models. Given a $g^E$-model, one can try to solve an equation system based on the isoactivity condition \citep{sorensen1979, swank1986, teh2002}. As this condition is only a necessary but not sufficient criterion for stable equilibrium phases, more equations and conditions have to be taken into consideration to ensure correct solutions \citep{sorensen1979}. Additionally, these kind of methods usually rely on a good initial guess for the resulting compositions of the phases to prevent convergence to trivial solutions \citep{sorensen1979, swank1986}. Thus, most $g^E$-model-based approaches employ the tangent plane criterion \citep{baker1982}, which is based on the minimization of Gibbs energy \citep{gibbs1873, gibbs1876}. For a given feed composition, the stable phases are therein found by an optimization algorithm. There exist various examples for such approaches. Listing just a few, they can be classified in deterministic \citep{michelsen1982_1, michelsen1982_2, mcdonald1995_1, mcdonald1995_2, mitsos2007, wasylkiewicz2013} or stochastic optimization algorithms \citep{rangaiah2001, bonilla-petriciolet2011}. We refer to the literature for a thorough summary \citep{teh2002, zhang2011, piro2016}. 

The present work provides a generalization of the convex envelope method (CEM) \citep{ryll2009, ryll2012}, which based on the tangent plane criterion \citep{baker1982}, but can also be seen as a surrogate model, as it constructs an approximation of the phase equilibrium within a discretized space. Contrary to the approaches mentioned beforehand, the CEM constructs a fluid phase equilibria diagram for the (discretized) composition space of a given mixture as a whole and stores it in a piecewise linear representation. Afterwards, phase splits can be computed in a robust and fast way for given feed compositions. The number of phases for a given feed is computed by the CEM and thus has not to be specified beforehand. Obviously, the construction for the whole composition space is computationally inefficient, if only a few feed compositions for fixed conditions (temperature and pressure) have to be evaluated. But it is advantageous in conceptual process design, when many different process options with possible recycles have to be evaluated in a robust way. Example applications are conceptual flowsheet evaluators \citep{ryll2009} or automated flowsheet synthesis \citep{yeomans1999, goettl2021}. In the CEM, the composition space is discretized into equally distributed points and at every point the Gibbs energy of mixing is calculated. Combined with those values, the discretization points represent a graph, which is used as a basis to construct a convex envelope around it. Given this surface, the tangent plane criterion \citep{baker1982} is used to determine all stable equilibrium states in the whole composition space. \cite{ryll2009} and \cite{ryll2012} presented the CEM to determine phase equilibrium diagrams in the liquid phase for up to four components. While it is clear from a thermodynamic point of view, and also stated in \citep{ryll2009, ryll2012}, that the CEM is applicable to systems with an arbitrary number of components, such a generalization is still missing. This generalization will be provided by the present work, which presents a changed version of the CEM. The novel CEM method is shown to work for several systems with three, four, five, and six components presented in the literature.

\section{Methodology}
\label{methodology}

\subsection{General Idea}
\label{general_idea}
\begin{figure}
	\includegraphics[width=0.9\linewidth]{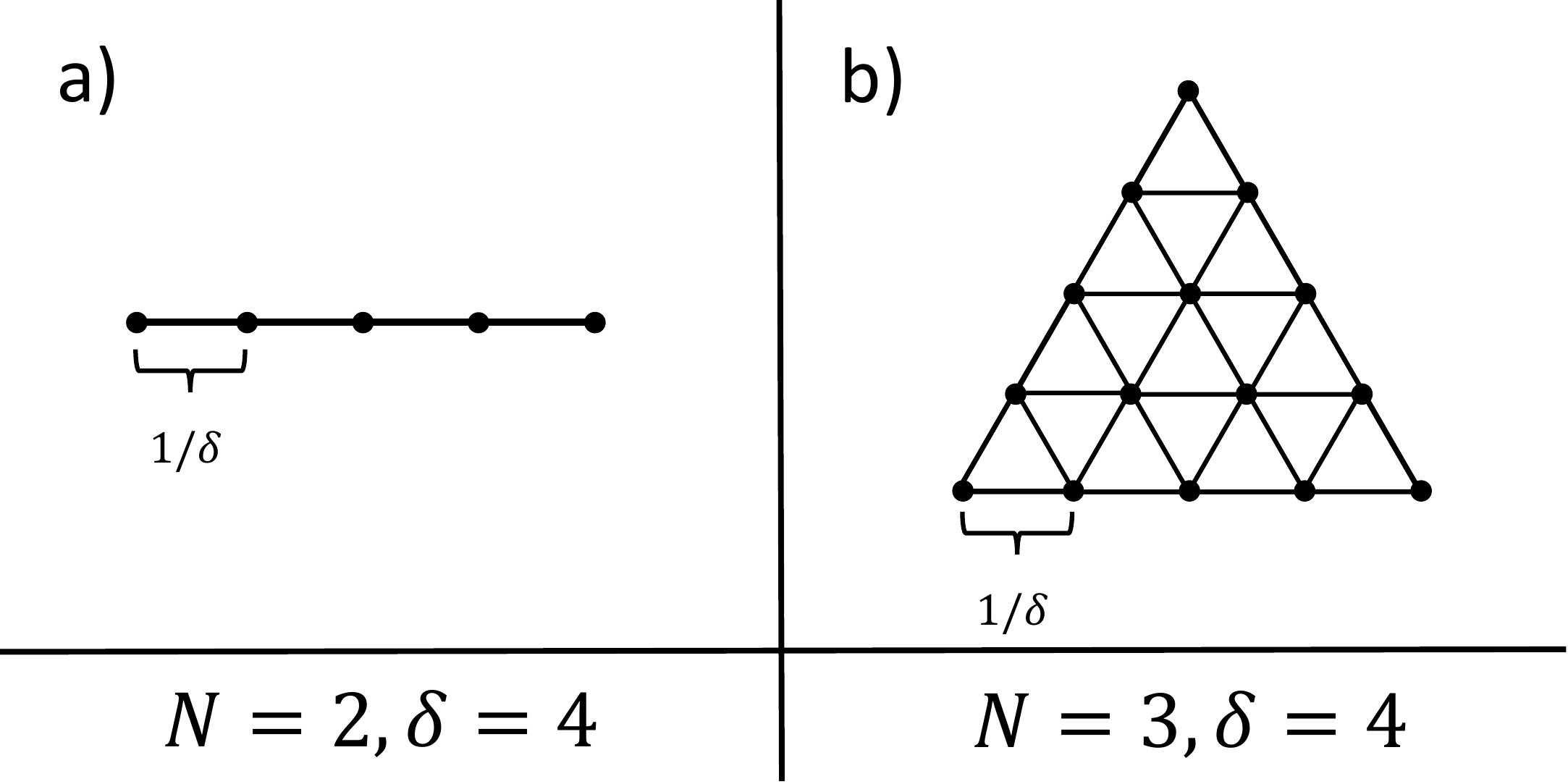}
	\caption{Example for the discretization of simplices depending on the parameter $\delta$ for systems with $N=2$ or $N=3$ components.}
	\label{discretization_figure}
\end{figure}

\begin{figure}
	\includegraphics[width=0.9\linewidth]{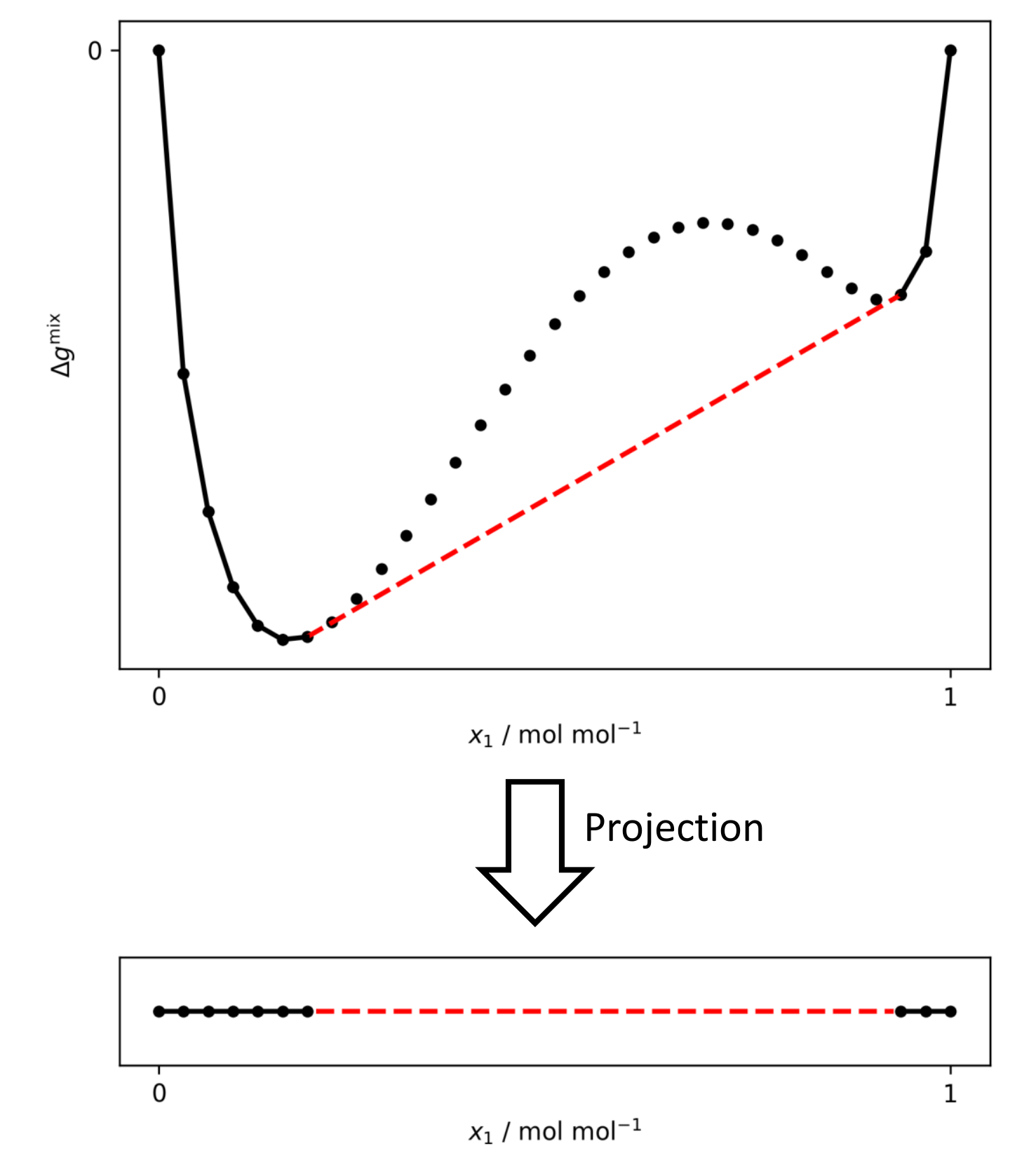}
	\caption{Example for the convex envelope of a $\Delta g^{\mathrm{mix}}$ graph for a binary system (solid black and dashed red line segments). The homogeneous simplices are the black, solid line segments (those simplices connect only points which are direct neighbours in the discretized composition space). The red, dashed line segment is a heterogeneous simplex (it spans over a multiphase region). The lower part shows the projection of the convex envelope to the discretized composition space.}
	\label{conv_hull_2}
\end{figure}

\begin{figure}
	\includegraphics[width=0.9\linewidth]{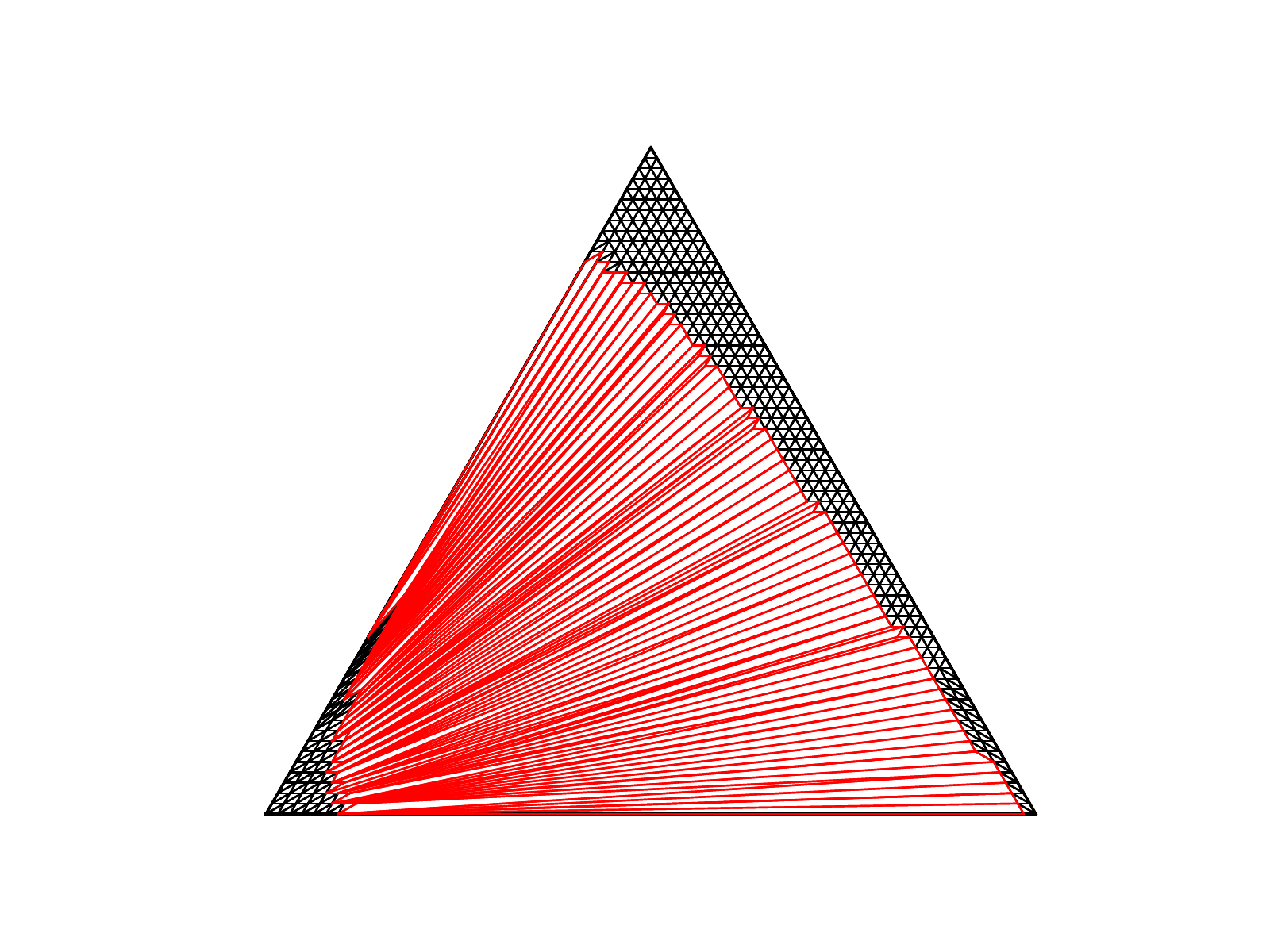}
	\caption{Example for the classification of homogeneous (black) and heterogeneous (red) simplices for a ternary system. The homogeneous simplices connect neighbouring points in the discretization space, while the heterogeneous simplices span over a larger area.}
	\label{after_projection_tern_space}
\end{figure}

\begin{figure}
	\includegraphics[width=0.9\linewidth]{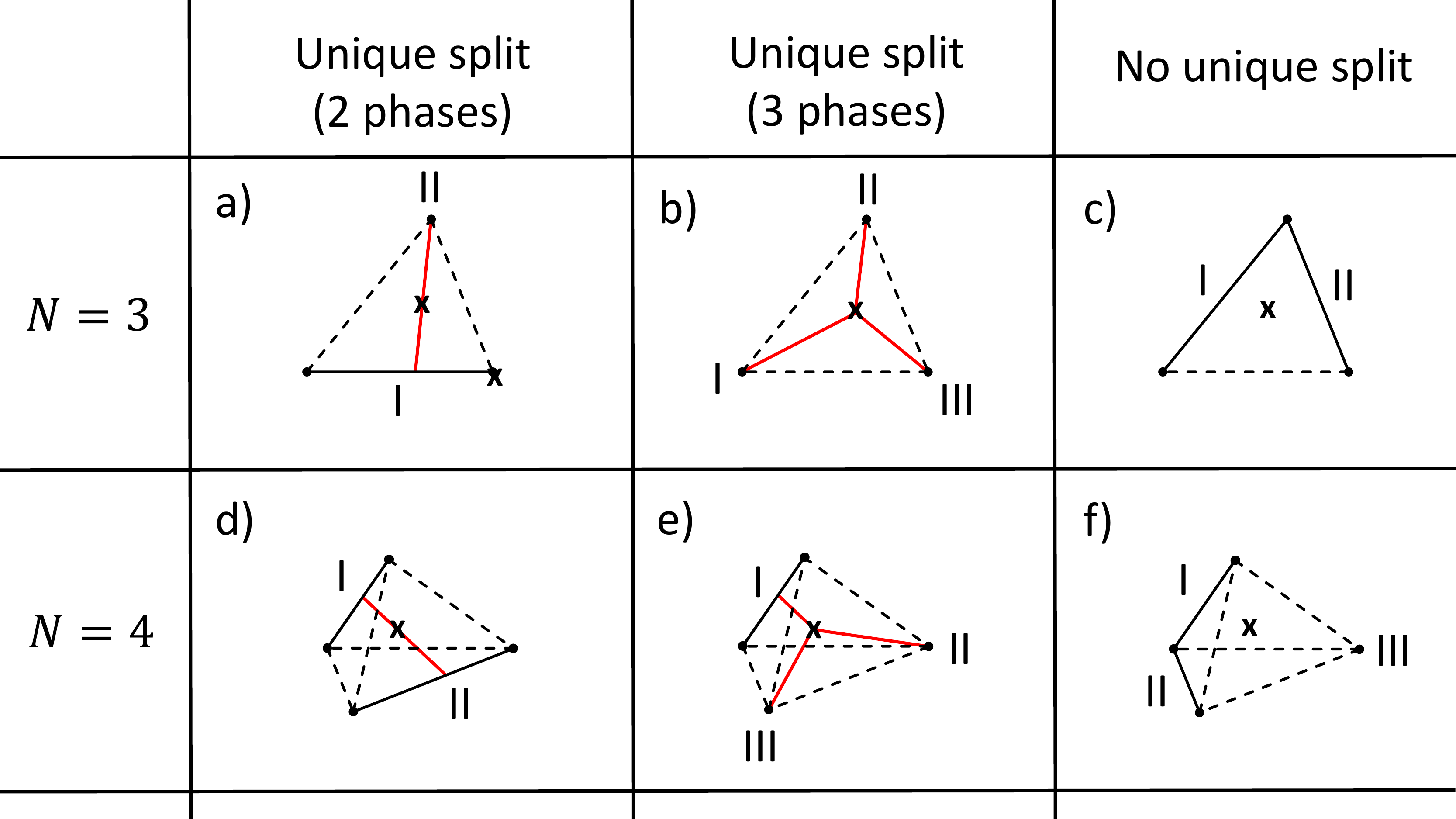}
	\caption{Examples for heterogeneous simplices. The homogeneous line segments are solid and the heterogeneous line segments are dashed. The considered feed compositions are marked with symbol x. The red lines in the first two columns show unique, linear splits of feeds inside the simplices. The last column shows two examples for simplices, which cannot be modelled in a unique way.}
	\label{simplex_example}
\end{figure}

We refer to \citep{ryll2009, ryll2012} for a detailed derivation and explanation of the CEM and just give a brief description of the general ideas for a system consisting of $N\in\mathbb{N}$ components:
\begin{itemize}
	\item [Step I)] Discretization of the composition space:\\
	A system consisting of $N$ components can be represented by a simplex with $N$ vertices. This simplex is discretized by choosing points inside it, which are uniformly distributed. This is done by specifying a minimal distance $\frac{1}{\delta}, \delta\in\mathbb{N}$ between the discretization points. A visualization of this concept is provided within Figure \ref{discretization_figure}. For $N=2$ and $\delta=4$, a binary system is discretized into 5 points with a minimal distance of $\frac{1}{4}$ between them (Figure \ref{discretization_figure} a)). For $N=3$ and $\delta=4$, the discretization yields 15 points in total (Figure \ref{discretization_figure} b)). Note that the number of points in the discretized space is bounded by $\mathcal{O}((\delta+1)^{N-1})$, but alongside Figure \ref{discretization_figure} it is also easy to visualize that for $N>2$, the number of points is less than $(\delta+1)^{N-1}$. For a detailed, mathematical explanation of the simplex representation and the discretization procedure we refer to the supplementary material.

	\item [Step II)] Determination of the $\Delta g^{\mathrm{mix}}$-graph and the convex envelope:\\
	For each point in the discretized space with molar fractions $\bm{x}$, the Gibbs energy of mixing \citep{rowlinson1982} is calculated:
	\begin{equation}
	\Delta g^{\mathrm{mix}}(\bm{x})=g(\bm{x})-\sum_{i=1}^{N}x_ig_i^{\mathrm{pure}}=\mathcal{R}T\sum_{i=1}^{N}\ln(x_i\gamma_i).
	\end{equation}
	The molar fractions $\bm{x}$ are transformed to cartesian coordinates $\bm{a}\in\mathbb{R}^n$ (described in the supplementary material) with $n=N-1$. Combined with the values for $\Delta g^{\mathrm{mix}}(\bm{x})$, one obtains points of the form $(\bm{a},g^{\mathrm{mix}}(\bm{x}))$, which represent a graph in $\mathbb{R}^{n+1}$. The convex envelope is constructed around the points of this graph.
	
	\item [Step III)] Classification of the simplices of the convex envelope:\\
	The convex envelope from Step II) consists of several simplices, i.e., boundary elements. For example, the simplices of the convex envelope in two dimensions are straight line segments, in three dimensions triangles. An example for the convex envelope of a binary system is shown in Figure \ref{conv_hull_2}. To obtain possible phase splits, one has to classify the simplices of the envelope into homogeneous (black, solid line segments) and heterogeneous (red, dashed line segment). A homogeneous simplex connects only points which are direct neighbours in the discretized composition space, i.e., points with minimal distance between each other, specified by the parameter $\delta$ (see also Figure \ref{discretization_figure}). All other simplices are heterogeneous and span over a multiphase region. As shown in the lower part of Figure \ref{conv_hull_2}, the simplices can be classified after projection of the convex envelope to the discretized composition space. The values for $\Delta g^{\mathrm{mix}}$ are discarded and the remaining distances determine, if two points are neighbouring or not. In a ternary system, the simplices are triangles consisting of straight line segments, where each line segment either connects two neighbouring points in the discretization space, i.e., a homogeneous line segment, or not, i.e., a heterogeneous line segment. Simplices, which connect only neighbouring points and thus consist only of homogeneous line segments are called homogeneous, all other simplices are called heterogeneous. Figure \ref{after_projection_tern_space} visualizes the classification into homogeneous (black) and heterogeneous (red) simplices for a ternary system.\\
	For each heterogeneous simplex, one needs to check, if it can be used to model a unique phase split in a linearized decanter model. This is shown alongside Figure \ref{simplex_example} in a similar manner as in \citep{ryll2009}: The homogeneous line segments are solid and the heterogeneous line segments are dashed. Figure \ref{simplex_example} a) shows a split of a feed in a ternary system into two phases: a unique straight line through the feed, which ends in the compositions of the two phases. Figure \ref{simplex_example} b) shows a split of a feed in a ternary system into three phases (a unique plane, which contains the feed and the compositions of the three phases). In Figure \ref{simplex_example} c), there are infinitely many possibilities to draw a straight line through the feed and the two phases. Therefore, we can not model this simplex uniquely and omit it for further analysis in Step IV). Figures \ref{simplex_example} d)-f) show the same situations for systems with four components.\\
	As mentioned in \citep{ryll2009}, heterogeneous simplices, which are not feasible for a unique, linear phase split, occur rarely and only at the boundary of multiphase regions, i.e., at locations where heterogeneous and homogeneous simplices are direct neighbours. They nearly vanish when more points are specified within the discretization. We provide detailed results on the effect of a low value for $\delta$ on the accuracy of the CEM later on.	 
\end{itemize}

Steps I)-III) yield the phase diagram for the given system at the temperature used to calculate $\Delta g^{\mathrm{mix}}$. Examples are shown in the Results section. These steps have to be done only once per system and temperature.

\begin{itemize}
	\item [Step IV)] Computation of phase splits:\\
	For a given feed composition, one has to check if the point lies within a feasible, heterogeneous simplex, which has been stored at Step III). The occurring phase split is computed, depending on the simplex' structure and the lever arm rule. This step is explained in detail in the next subsection.
\end{itemize}

Step I) and II) can be executed for an arbitrary number of components without a change. At Step III) one has to classify the simplices into homogeneous and heterogeneous. \cite{ryll2009} have provided a graphical evaluation scheme for this task that distinguishes several cases and does the decision on a case-by-case basis. The scheme is only presented for up to four components. An extension to more components would be very cumbersome and also hard to grasp. The number of different cases for the heterogeneous simplices increases rapidly (binary: 1, ternary: 3, quaternary: 8). This problem also remains present at Step IV) as one has to define the type of linear split for every possible simplex topology. To address these challenges, we propose a mathematical framework for the classification of the heterogeneous simplices that is general and allows the computation of splits for an arbitrary number of components and phases.

\subsection{Mathematical Framework}
We consider a system consisting of $N\in\mathbb{N}$ components, which is represented by a $n$-simplex $U$ in $\mathbb{R}^n$ with $n=N-1$ (e.g., a 3-component system is represented by a triangle in $\mathbb{R}^2$, i.e., a 2-simplex). Given such an $n$-simplex, we now want to switch easily between cartesian coordinates (i.e., coordinates in $\mathbb{R}^n$) and molar fractions. This can be done by using barycentric coordinates. Barycentric coordinates of a point inside a simplex are coordinates with respect to the vertices of this simplex, which are non-negative and sum up to 1. As will be seen later on, barycentric coordinates of a point inside a simplex, which represents the whole composition space, are identical with molar fractions of this point in the composition space. For a detailed explanation of the simplex geometry, which is used throughout this work we refer to the supplementary material. Note that the transformation by barycentric coordinates is linear and thus the following geometric results should work also directly in the molar fraction space. We transform the coordinates mainly for the sake of illustration, as we believe some concepts are easier to visualize this way. If not stated otherwise, we refer from now on to cartesian coordinates. We assume that the system has been discretized and that the convex envelope of the $\Delta g^{\mathrm{mix}}$-graph has been computed. We start with defining heterogeneous simplices:

\begin{defn}[Heterogeneous simplex]
	Consider a system consisting of $N$ components, represented by a $n$-simplex $U$ with $n=N-1$, which was discretized with parameter $\delta$. Let $\bm{h}_1, \dots, \bm{h}_N\in\mathbb{R}^n$ be points in the discretized space ($\bm{h}_1, \dots, \bm{h}_N$ are the cartesian coordinates after transformation of the molar fractions), which define a simplex $H=\text{conv}(\{\bm{h}_1, \dots, \bm{h}_{N}\})$, originating from Step II) in subsection \ref{general_idea}. We call $H$ \textit{heterogeneous} or a \textit{heterogeneous simplex} if there exist at least two indices $i, j\in\{1, \dots, N\}$ with $i\neq j$ so that $\bm{h}_i$ and $\bm{h}_j$ are not neighbouring points in the underlying discretization of the composition space. Neighbouring points are defined by having a minimal distance dependent on $\delta$ in the discretized space.
\end{defn}

From Step III) in subsection \ref{general_idea} we know that not every heterogeneous simplex is necessarily feasible for a unique, linearized decanter model. The following definition provides us with a subset of heterogeneous simplices, which can modelled in the desired way (this will be proven at the end of this subsection).

\begin{defn}[Phase block, isolated simplex]
	\label{def_isolation}
	Let $H=\text{conv}(\{\bm{h}_1, \dots, \bm{h}_{N}\})$ be a heterogeneous simplex. A \textit{phase block} is a subset $B\subseteq\{\bm{h}_1, \dots, \bm{h}_N\}$ with the following properties:
	\begin{itemize}
		\item [I)] For all $\bm{b}_i, \bm{b}_j\in B$ with $\bm{b}_i\neq\bm{b}_j$, it holds that $\bm{b}_i$ and $\bm{b}_j$ are neighbouring points in the underlying discretized composition space.
		
		\item [II)] $B$ is maximal w.r.t. property I), i.e., for all $\bm h \in \{\bm{h}_1, \dots, \bm{h}_N\} \setminus B$ there is at least one $\bm b \in B$ such that $\bm h$ and $\bm b$ are not neighbouring points.
	\end{itemize}
	Note that a phase block $B = \{\bm{b}_1, \dots, \bm{b}_{k+1}\}$ defines a $k$-simplex $\text{conv}(B)$ which only consists of neighbouring points.\\
	A phase block $B$ is called \textit{isolated}, if for all $\bm h \in \{\bm{h}_1, \dots, \bm{h}_N\} \setminus B$ and all $\bm b \in B$ it holds that $\bm h$ and $\bm b$ are not neighbouring points. Note that this is a stricter requirement than property II).\\
	A heterogeneous simplex $H$ is called \textit{isolated}, if all phase blocks in $H$ are isolated.
\end{defn}

\begin{rema}[Decomposition property of heterogeneous, isolated simplices]
	\label{decomposition_rema}
	It follows directly from the definition that any isolated heterogenous $(N-1)$-simplex $H = \text{conv}(\{\bm{h}_1, \dots, \bm{h}_N\})$ decomposes uniquely into $m \leq N$ isolated phase blocks $B_1, \dots, B_m$ such that $\bigcup_{i = 1}^m B_i = \{\bm{h}_1, \dots, \bm{h}_N\}$. Furthermore, $B_i \cap B_j = \emptyset$ for $i \neq j$, and also $\text{conv}(B_i) \cap \text{conv}(B_j) = \emptyset$ as neighbouring points have minimal distance in the disrectized composition space.
\end{rema}

These definitions can be visualized using Figure \ref{simplex_example}: the simplices in a)-f) are all heterogeneous, since all of them contain at least one heterogeneous line segment. Examples for phase blocks are the line segment I (a $1$-simplex) and the point II (a $0$-simplex) in part a) of Figure \ref{simplex_example} (these are also isolated, as the only connections to other phase blocks are heterogeneous line segments). Figure \ref{simplex_example} part c) shows two phase blocks (the line segments I and II), which are not isolated as they are connected. Also, those two line segments do not just form one phase block, since a third homogeneous line segment would be needed to represent a $2$-simplex (i.e., a triangle). Note that one could also spare the classification into homogeneous and heterogeneous simplices, as a homogeneous simplex can be seen as a simplex consisting of just one phase block. We keep the distinction into two types of simplices though to emphasize the differences: a heterogeneous simplex, contrary to a homogeneous simplex, models a phase split.

Additionally, note that in Figure \ref{simplex_example}, the simplices, which can be modelled uniquely, are exactly the ones that are isolated. Before we can prove that the isolated, heterogeneous simplices are exactly the ones, which can be modelled within a unique, linearized decanter model, we need to describe a phase split inside a heterogeneous simplex:

\begin{defn}[Phase split simplex]
	\label{splitsimplex}
	Let $H = \text{conv}(\{\bm{h}_1, \dots, \bm{h}_N\}) \subset \mathbb R^n$ be a heterogeneous, isolated simplex, with isolated phase blocks $B_1, \dots, B_m$. Let $\bm{x}$ be the molar fractions of a feed with cartesian coordinates $\bm{a}\in\mathbb{R}^n$ so that $\bm{a}$ lies in $H$. A split of $\bm a$ into $m$ phases is defined by a $(m-1)$-simplex $S = \text{conv}(\{\bm{p}_1, \dots, \bm{p}_m\}) \subseteq H$ with vertices $\bm{p}_1, \dots, \bm{p}_m\in\mathbb{R}^n$ so that the following properties are fulfilled:
	\begin{itemize}
		\item [I)] We have  $\bm{a} \in S$.
		
		\item [II)] For all $i\in\{1, \dots, m\}$, we have $\bm{p}_i \in \text{conv}(B_i)$.
	\end{itemize}
	We call $S$ a \textit{phase split simplex with respect to $\bm a$}. The vertices of $S$ define the compositions of the resulting phases and we call $\bm p_i$ a \textit{phase}. The cartesian coordinates of the vertices $\bm{p}_1, \dots, \bm{p}_m$ can be transformed to molar fractions (i.e., the compositions of the resulting phases) via the barycentric coordinates with respect to the simplex which represents the whole $N$-component system. Furthermore, the split ratios $\lambda_1, \dots, \lambda_m$ are given exactly by the barycentric coordinates of $\bm{a}$ with respect to the vertices of $S$.
\end{defn}

For example, a phase split simplex for a given feed for two phases is a $1$-simplex and therefore a straight line, which contains the feed and is defined by the compositions of the resulting phases. The barycentric coordinates of the feed with respect to that line yield exactly the well known lever arm rule (for barycentric coordinates $\lambda_1, \dots, \lambda_m$ holds $\lambda_i\geq 0$ and $\sum_{i=1}^{m}\lambda_i=1$).\\

Now consider an arbitrary feed composition within an isolated, heterogeneous simplex. This means that the simplex spans over a multiphase region and we want to compute the compositions of the resulting phase split. The following theorem proves on the one hand that for every feed composition within an isolated, heterogeneous simplex, there exists a unique linear phase split (i.e., a phase split simplex from Definition \ref{splitsimplex}). On the other hand it also provides a formula for the computation of the compositions of the resulting phases.

\begin{theo}
	\label{main_theorem}
	Let $\bm{x}$ be a feed composition and $\bm{a} \in \mathbb R^n$ its representation in cartesian coordinates, located in an isolated, heterogeneous $(N-1)$-simplex $H = \text{conv}(\{\bm{h}_1, \dots, \bm{h}_N\})$, with $k>1$ phase blocks $B_1, \dots, B_k$. For any phase block $B_i$, denote by $\Lambda_i=\{j\in\{1, \dots, N\}\ \vert \ \bm{h}_j \in B_i \}$ its corresponding index set. Let $(\tilde{\lambda}_1, \dots, \tilde{\lambda}_N)$ be the barycentric coordinates of $\bm{a}$ with respect to $H$. Furthermore, set $\lambda_i := \sum_{j \in \Lambda_i}\tilde \lambda_j$. \\\\
	Then $S = \text{conv}(\{\bm{p}_1, \dots, \bm{p}_k\})$ with $\bm{p}_i = \sum_{j\in\Lambda_i}\alpha_j\bm{h}_j$, where $\alpha_j = \frac{\tilde{\lambda}_j}{\lambda_i}$, is a unique phase split simplex for $\bm a$. The split ratios for the phases are given by $(\lambda_1, \dots, \lambda_k)$.\\
	For a proof, we refer to the supplementary material.
\end{theo}

With Theorem \ref{main_theorem}, we know how to model unique, linear phase splits in isolated, heterogeneous simplices. Therefore, we propose the following modifications to Step III) and IV) in subsection \ref{general_idea}: 

\begin{itemize}
	\item [Step III)] Classification of the simplices of the convex envelope:\\
	We check for all heterogeneous simplices, if those are isolated and if so process those to Step IV). It is easy to check that for binary, ternary, and quaternary systems this classification does not change the methodology described in \citep{ryll2009, ryll2012}.
	\item [Step IV)] Computation of phase splits:\\
	For a given feed with molar fractions $\bm{x}$, we transform $\bm{x}$ to cartesian coordinates $\bm{a}$ and check if $\bm{a}$ lies within an isolated, heterogeneous simplex $H$ (with $k$ phase blocks). If this is the case, we use Theorem \ref{main_theorem} to obtain the split ratios $(\lambda_1, \dots, \lambda_k)$ and the cartesian coordinates $\bm{p}_1, \dots, \bm{p}_k$ of the resulting phases. Using the inverse transformation of coordinates that was used to obtain $\bm{a}$ from $\bm{x}$, one can get molar fractions corresponding to each phase.
\end{itemize}

\subsection{Implementation}
The whole framework was implemented with the programming language Python (version 3.9). The convex envelope is found by usage of the QuickHull-algorithm \citep{barber1996}, which is provided by the Qhull library within the package scipy \citep{virtanen2020}. An implementation, which allows execution of Step III) from subsection \ref{general_idea} parallelized, is published via GitHub: \url{https://github.com/grimmlab/cem}. All results were generated on a machine with an AMD EPYC 7542 32-core (64 threads) processor and 512 GB of RAM.

\section{Results}
\label{results}
\subsection{Qualitative Evaluation}
\begin{figure}
\includegraphics[width=0.9\linewidth]{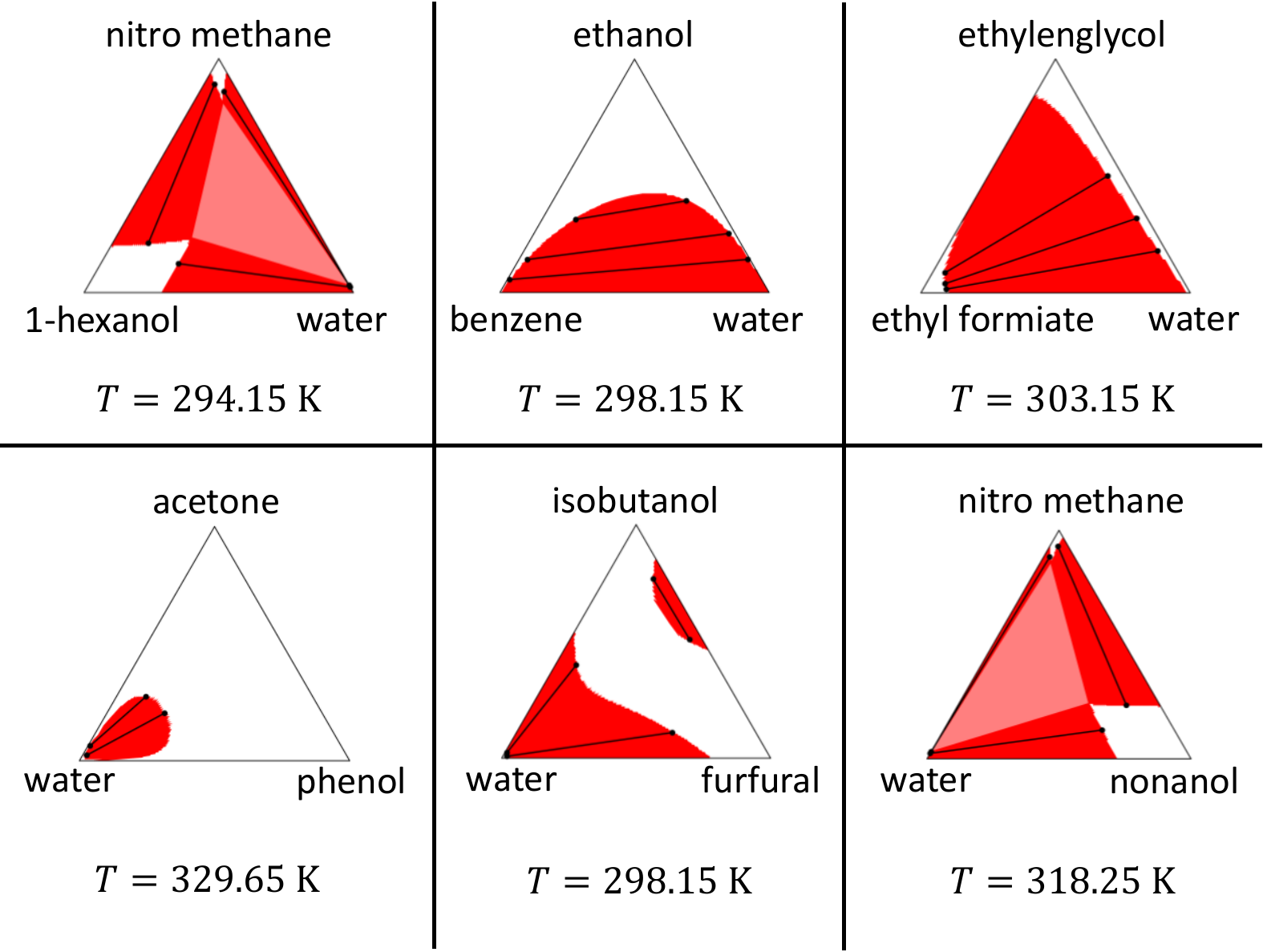}
\caption{Selection of ternary systems with UNIQUAC parameters from \citep{ryll2009, ryll2012} constructed with the generalized CEM at atmospheric pressure. The plots display molar fractions. The transparent red areas in the systems 1-hexanol -- nitro methane -- water and water -- nitro methane -- nonanol display three phase regions, all other red areas display two phase regions (for every two phase region, a few example tie lines are plotted in black).}
\label{ryll_tern}
\end{figure}

The CEM was already shown to work for ternary and quaternary systems by \citep{ryll2009, ryll2012}. Therefore, we show first that the generalized version of the CEM works in the same way for systems, which were examined there. As no numerical data for comparison (feed streams and resulting phases) is given in \citep{ryll2009, ryll2012}, we show ternary diagrams constructed by the generalized CEM for a selection of systems from \citep{ryll2009, ryll2012} using the same binary parameters for the UNIQUAC model \citep{prausnitz1999}. It is quite difficult to display the phase equilibrium diagram for a quaternary system in a way that one can actually get useful information out of it, let alone comparing it to other phase diagrams. Therefore, we omit a graphical evaluation of the quaternary systems shown in \citep{ryll2009, ryll2012} and show later on in a numerical way that our approach works for systems with more than three components. 
Figure \ref{ryll_tern} shows six ternary systems reported in \citep{ryll2009, ryll2012} with various types of liquid phase equilibria. Note that the systems 1-hexanol -- nitro methane -- water and water -- nitro methane -- nonanol also show a three liquid phase region, which is enclosed by the two phase regions. When comparing Figure \ref{ryll_tern} to the ternary plots shown in \citep{ryll2009, ryll2012}, one can see that the displayed phase equilibria are the same.

\subsection{Quantitative Evaluation}
To evaluate our approach quantitatively, we will compare its results with experimental data from several sources \citep{chen2000, chen2001, yuan2018, yuan2019, yuan2020}. All of the listed authors provide binary interaction parameters for a $g^E$-model (either UNIQUAC or NRTL \citep{prausnitz1999}), experimental data on occurring phase splits, and the resulting root-mean-square-error (RMSE) of the model compared to experimental data. Unfortunately, the data points generated by the models were not reported. But as the RMSE is rather low ($<0.01$ for almost all directly fitted examples in \citep{chen2000, chen2001, yuan2018, yuan2019, yuan2020}), we compare the CEM to the reported experimental data as a work-around. We measure the accuracy of the generalized CEM by mean deviation (MD) to given data:
\begin{equation}
\textmd{MD}=\frac{\sum_{i=1}^{N}\sum_{j=1}^{M}\sum_{k=1}^{P}|x_{ijk}^{\text{source}}-x_{ijk}^{\text{CEM}}|}{NMP}.
\end{equation}
$N$ describes the number of components, $M$ the number of examined feed streams and $P$ the number of phases (in \citep{chen2000, chen2001, yuan2018, yuan2019, yuan2020} only splits into two phases were reported). $x^{\text{source}}$ refers to the molar fraction reported in the literature. 

Detailed results for systems, which were reported in \citep{chen2000, chen2001, yuan2018, yuan2019, yuan2020}, containing three, four, five, and six components are provided in the supplementary material. For all examined systems, the CEM was able to calculate the occurring phase splits with high accuracy. Additionally, we report detailed information on the computation times of the CEM in the supplementary material. For ternary systems, our implementation of the CEM constructs phase equilibria in up to 2 seconds with serial execution of Step III) from subsection \ref{general_idea}. For systems with more components, Step III) from subsection \ref{general_idea} was executed parallelized (the respective settings for parallelization are published alongside the code on GitHub). For quaternary systems, computation time per system is below 1 minute, whereas for systems containing five components, up to 3 minutes are required. For systems containing six components, it takes up to 50 minutes per system to construct the phase equilibrium.

\subsection{Analysis of the impact of the discretization parameter $\delta$}
\begin{figure}
	\includegraphics[width=0.9\linewidth]{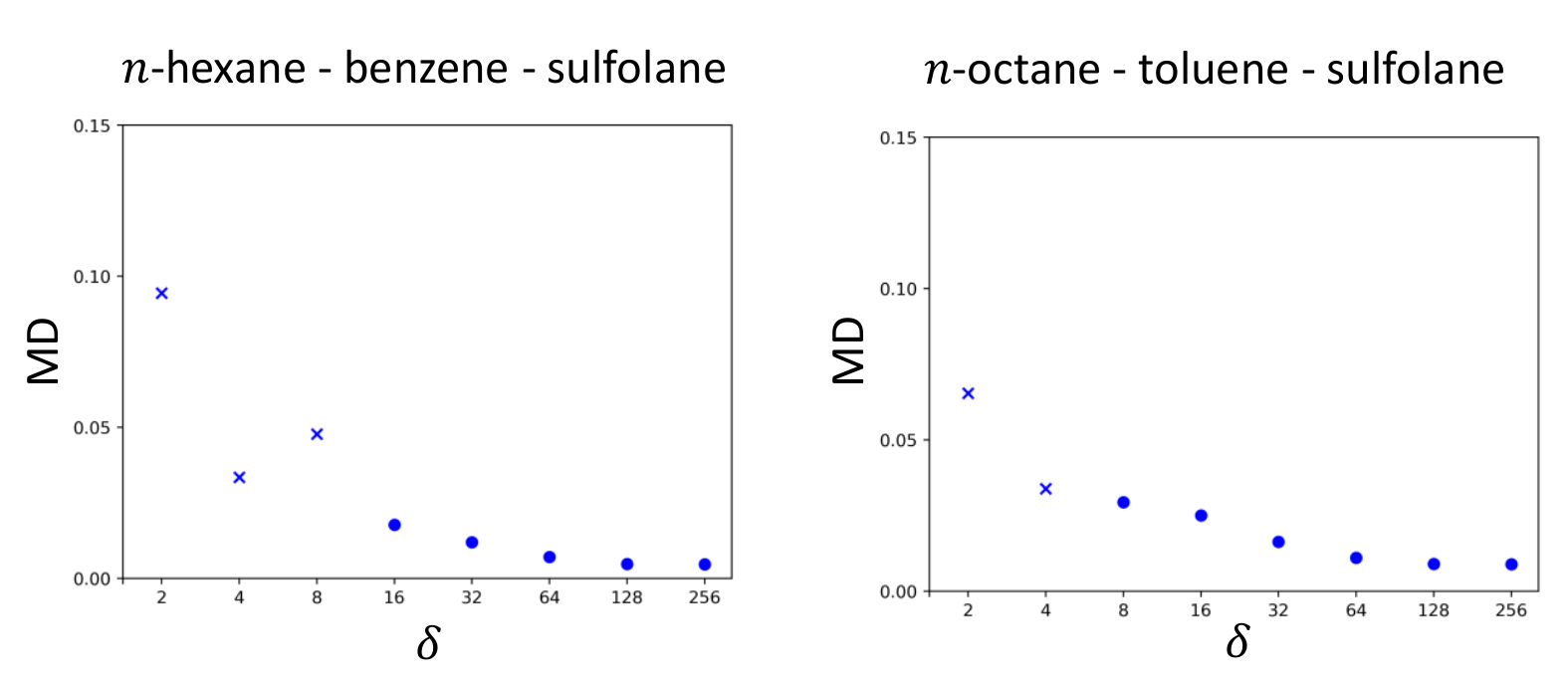}
	\caption{MD values for several choices for $\delta$ for the systems $n$-hexane -- benzene -- sulfolane and $n$-octane -- toluene -- sulfolane. MD was calculated using phase split data and parameters from \citep{chen2000}. Some points in the graph are marked with the symbol x, which indicates that no phase split was found for some given feed stream compositions for this choice of $\delta$.}
	\label{delta_analysis_grafik}
\end{figure}

\begin{figure}
	\includegraphics[width=0.9\linewidth]{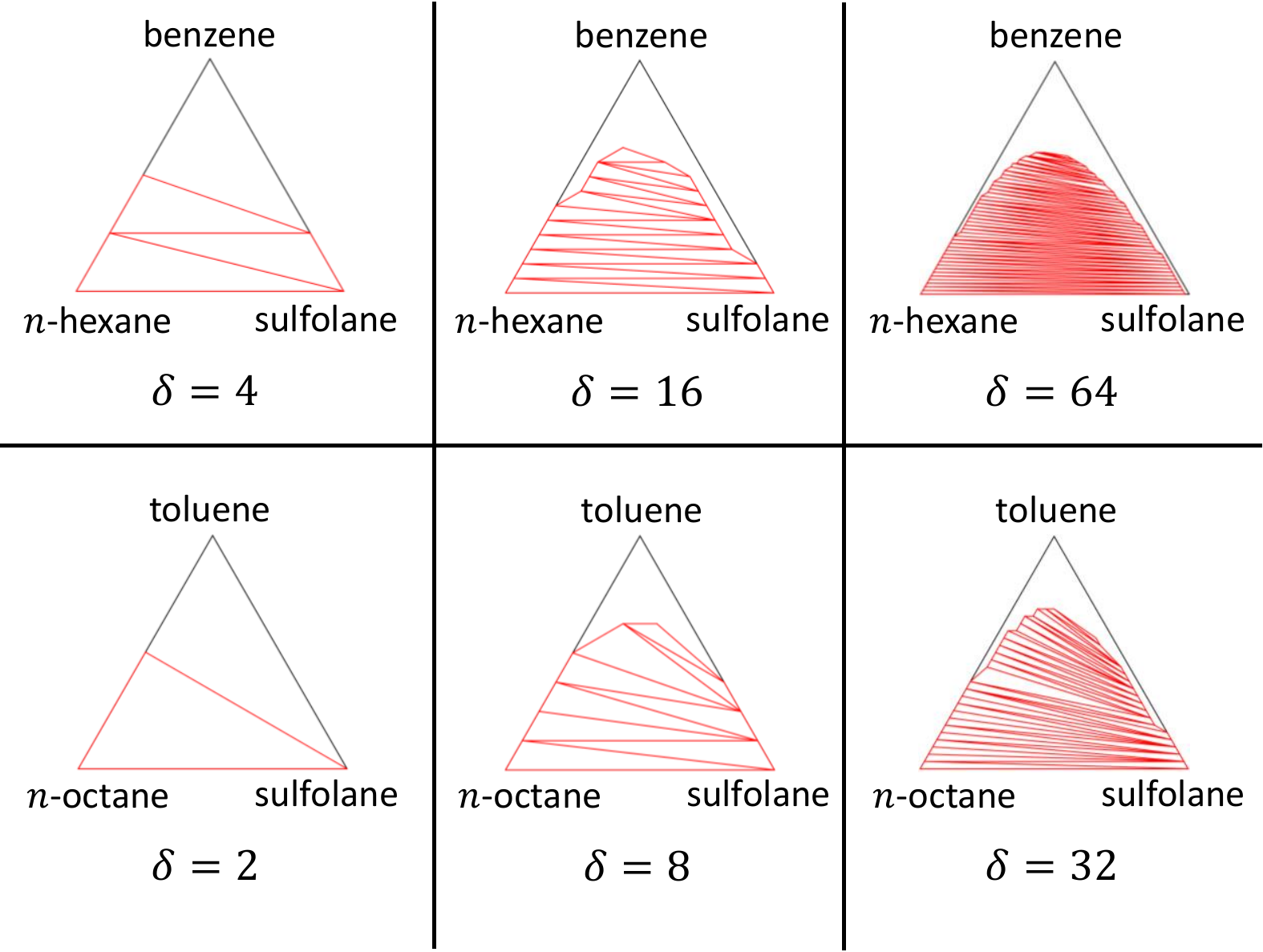}
	\caption{Variation of $\delta$ for the construction of ternary diagrams for the systems $n$-hexane -- benzene -- sulfolane and $n$-octane -- toluene -- sulfolane from \citep{chen2000} at 298.15 K and atmospheric pressure. The red triangles are the heterogeneous simplices (i.e., simplices, which span over a phase split region).}
	\label{delta_analysis_tern_sys}
\end{figure}

To show the impact of the discretization parameter $\delta$ on the accuracy of the calculated phase splits, we plot the MD for two ternary example systems from \citep{chen2000} for various choices of $\delta$ in Figure \ref{delta_analysis_grafik}. For low values of $\delta$, the CEM is not always able to determine that for a given feed stream composition there will be a phase split according to the data. These points in Figure \ref{delta_analysis_grafik} are marked with the symbol x. But as expected, if $\delta$ is increased, the MD decreases and the CEM is able to calculate phase splits for all given data points correctly. Also it can be seen that already for rather low values of $\delta$ as for example 16, an accurate calculation of phase splits is possible. The residual deviation for large $\delta$ is originating from the deviation of the model and the experiments, with which we compare us. Figure \ref{delta_analysis_tern_sys} shows the heterogenous simplices (in red) for different choices of $\delta$. As can be seen, when $\delta$ is increased, the boundary of the two phase region gets smoother and approximates the shape defined by the non-discretized model.

\clearpage
 
\section{Discussion}
The present work shows a generalization of the CEM \citep{ryll2009, ryll2012}, for systems with an arbitrary number of components. Liquid phase equilibria are obtained by discretization of the whole composition space and construction of the convex envelope for the Gibbs energy of mixing graph. A theoretical framework for the extension to an arbitrary number of components is given and the approach was proven to work for systems from the literature with up to six components.

While Theorem \ref{main_theorem} gives the possibility to compute phase splits for systems with an arbitrary number of components, computational complexity becomes a limiting factor, when systems with more than six components are examined. Particularly the construction of the convex envelope is a challenging task. For this, we utilize the package scipy \citep{virtanen2020}, which bases the construction of the convex envelope on the QuickHull-algorithm \citep{barber1996} provided by the Qhull library. In the Qhull library documentation it is mentioned that complexity increases rapidly with the number of input points (defined by the discretization of the composition space) and also with the dimension of the input (defined by the number of components). For the CEM, we not only require the points that form the convex envelope, but also the connections between those points, i.e. the facets. According to the Qhull library documentation, this requires a lot of virtual memory and finally leads to a rapid decrease of performance. To approach higher order systems, a effective implementation of a parallelized convex envelope algorithm could be a solution to this problem. Theoretical research focusing on the construction of convex envelopes in a parallelized fashion is for example provided by \cite{amato1994} and \cite{blelloch2020}, but there is no implementation available, which could be integrated into our framework easily, to compare the resulting performances. Additionally, it is not clear, if those algorithms are able to construct the convex envelope with less usage of virtual memory. There exist other algorithms, which are able to solve certain problem classes faster than the Qhull framework \citep{chadnov2004}, e.g., two-dimensional input and special topologies. But to our knowledge there exists no implementation, which exceeds the Qhull framework in terms of generality and effectiveness, as it can handle an arbitrary number of dimensions. Also the classification of the heterogeneous simplices (Step III) in \ref{general_idea}) is a time consuming step, but contrary to the construction of the convex envelope, it can be parallelized quite easily. The discretization of the composition space and the computation of $\Delta g^{\mathrm{mix}}$-graph do not need a lot of computation time compared to the former mentioned steps and are negligible.

The presented method constructs phase equilibria in a robust way for the whole composition space for fixed temperature and pressure. Obviously, the CEM may not be a useful tool for optimizing temperature in a decantation process. But it is advantageous, if a lot of different feed stream compositions at fixed conditions are examined, e.g., for the evaluation of a large number of feed streams in the conceptual design process.

In the present work, only liquid phase equilibria are constructed. But as already mentioned in \citep{ryll2009, ryll2012}, the CEM could additionally be applied for the calculation of arbitrary phase equilibria (e.g. vapor-liquid and solid-liquid). Integration of this feature into the present framework is out of the scope of this work, but an interesting option for future research.

\section{Conclusion}
This work presents a mathematical framework for the calculation of liquid phase equilibria for systems with an arbitrary number of components. The presented approach is a generalization of a method, which has already been shown to work for systems with up to four components. The generalized approach is shown to work alongside various examples from the literature with up to six components. Application to higher order systems is mainly limited by computational complexity of the construction of the convex envelope. Further research on parallelization algorithms and effective usage of virtual memory would be needed, to overcome this issue in the future. Another direction for further research could be the integration of vapor-liquid or solid-liquid equilibria into the framework.

\section*{Acknowledgement}
This work was funded by the Deutsche Forschungsgemeinschaft (DFG, German Research Foundation)
– 466387255 – within the Priority Programme ”SPP 2331: Machine Learning in Chemical
Engineering”.

\bibliographystyle{elsarticle-harv} 
\bibliography{goettletal2023}






\clearpage

\setcounter{section}{0}
\renewcommand{\thesection}{S\Roman{section}}
\setcounter{equation}{0}
\setcounter{figure}{0}
\setcounter{table}{0}
\setcounter{defn}{0}
\setcounter{theo}{0}
\setcounter{rema}{0}
\setcounter{lemm}{0}
\renewcommand{\theequation}{S\arabic{equation}}
\renewcommand{\thedefn}{S\arabic{defn}}
\renewcommand{\thetheo}{S\arabic{theo}}
\renewcommand{\therema}{S\arabic{rema}}
\renewcommand{\thelemm}{S\arabic{lemm}}

\begin{center}
	\textbf{\large Supplementary Material: Convex Envelope Method for determining liquid multi-phase equilibria in systems with arbitrary number of components}
\end{center}

\section{Mathematics}
\subsection{Simplex geometry}
\label{appendix_simplex_geometry}
\begin{defn}[Convex envelope/hull]
	Let $k, n\in\mathbb{N}$ and $\bm{u}_1, \dots, \bm{u}_{k}\in\mathbb{R}^n$. The \textit{convex envelope} or \textit{convex hull} of $\bm{u}_1, \dots, \bm{u}_{k}$ is the smallest convex set in $\mathbb{R}^n$, which contains $\bm{u}_1, \dots, \bm{u}_{k}$. We refer to it with $\text{conv}(\{\bm{u}_1, \dots, \bm{u}_{k}\})$.
\end{defn}

\begin{defn}[$k$-simplex]
	Let $k, n\in\mathbb{N}$ with $k\leq n$ and $\bm{u}_1, \dots, \bm{u}_{k+1}\in\mathbb{R}^n$ so that $\bm{u}_1-\bm{u}_{k+1}, \dots, \bm{u}_k-\bm{u}_{k+1}$ are linearly independent. A $k$-\textit{simplex} $U$ with vertices $\bm{u}_1, \dots, \bm{u}_{k+1}$ is defined as the convex envelope $\text{conv}(\{\bm{u}_1, \dots, \bm{u}_{k+1}\})$ of the points $\bm{u}_1, \dots, \bm{u}_{k+1}$. We often say $U$ is represented by $\bm{u}_1, \dots, \bm{u}_{k+1}$.
\end{defn} 

\begin{rema}[Simplex representation of $N$-component systems]
	A system consisting of $N\in\mathbb{N}$ components can always be interpreted as a $n$-simplex in $\mathbb{R}^n$ with $n=N-1$. For example a 3-component system can be visualized in $\mathbb{R}^2$ as a 2-simplex (e.g., an equilateral triangle). When mixtures in a $N$-component system are examined, usually molar fractions are used (e.g., to calculate thermodynamic properties). To be able to use results from geometry inside the simplex representing the system, one has to transform between molar fractions and cartesian coordinates. This can be done by the usage of barycentric coordinates (for a proof see for example \citep{rockafellar1970}).
\end{rema}

\begin{lemm}[Barycentric coordinates]
	\label{bary_theo}
	Let $k, n\in\mathbb{N}$ with $k\leq n$ and $\bm{u}_1, \dots, \bm{u}_{k+1}\in\mathbb{R}^n$ be the vertices of a $k$-simplex $U$. For every point $\bm{a}\in U$ there exist unique $\lambda_1, \dots, \lambda_{k+1}\in[0,1]$ so that 
	\begin{equation}\label{bary}
		\bm{a}=\sum_{i=1}^{k+1}\lambda_i\bm{u}_i \textrm{ and }\sum_{i=1}^{k+1}\lambda_i=1.
	\end{equation}
	$\lambda_1, \dots, \lambda_{k+1}$ are called \textit{barycentric coordinates} of $\bm{a}$ with respect to $U$.  
\end{lemm}

\begin{rema}
	Consider a $N$-component system, represented by a $n$-simplex $U$ with $n=N-1$. Barycentric coordinates $\lambda_1, \dots, \lambda_{N}$ of a point $\bm{a}\in U$ just describe the position of $\bm{a}$ with respect to the simplex' vertices. For a possible mixture inside the system, described by molar fractions $(x_1, \dots, x_N)$, it holds
	\begin{equation}
		\sum_{i=1}^{N}x_i=1 
	\end{equation} 
	and $x_i\geq0$ for all $i=1,\dots,N$. Because of the uniqueness statement in Lemma \ref{bary_theo}, it follows immediately that the molar fractions $(x_1, \dots, x_N)$ are the barycentric coordinates of a point $\bm{a}\in U$. This yields a possibility for transformation between molar fractions and cartesian coordinates by writing Equation \ref{bary} in matrix form (and computing the left-inverse of the left matrix).
\end{rema}

\subsection{Discretization of the composition space}
\label{appendix_discretization}
\begin{rema}
	A binary system can be discretized by specifying a minimal distance $1/\delta$, with $\delta\in\mathbb{N}$, for neighbouring points. The whole discretization space then contains $\delta+1$ points. In general, when a system with $N\in\mathbb{N}$ components is considered, we define the set of discretization points by
	\begin{equation}
		\mathcal{P}_\delta:=\{(\frac{p_1}{\delta}, \dots, \frac{p_N}{\delta})|p_i\in\mathbb{N}\text{ for }i=1,\dots, N\text{ and } \sum_{i=1}^{N}p_i=\delta\}.
	\end{equation}
	Note that the elements of $\mathcal{P}_\delta$ are the molar fractions $\bm{x}$ of the discretized points. As explained before, molar fractions $\bm{x}\in\mathcal{P}_\delta$ are identical to barycentric coordinates and thus can be transformed to cartesian coordinates with the help of Lemma \ref{bary_theo}. It is easy to see that every binary subsystem of $\mathcal{P}_\delta$ consists once again of $\delta+1$ points, but the total number of points increases exponentially with $N$. When we want to apply the CEM to a system with $N$ components, we construct a $n$-simplex ($n=N-1$) with unitary edge length, which represents the whole composition space. This space then is discretized as explained above.
\end{rema}

\subsection{Proof of Main Theorem}
\label{proof_appendix}
\begin{proof}
	We first show that $S$ is a phase split simplex for $\bm a$. We certainly have $\sum_{j \in \Lambda_i}\alpha_j = 1$ by definition and $\alpha_j \geq 0$ as $\tilde \lambda_j \geq 0$. Hence $\bm p_i \in \text{conv}(B_i)$, and we only need to show that $\bm{a} \in S$. For this, notice that $\lambda_i \geq 0$ and
	$$
	\bm a = \sum_{j = 1}^N \tilde \lambda_j \bm h_j \overset{(*)}{=} \sum_{i=1}^{k}\sum_{j\in \Lambda_i}\tilde \lambda_j \bm h_j = \sum_{i = 1}^k \lambda_i \cdot \sum_{j \in \lambda_i}\frac{\tilde \lambda_j}{\lambda_i} \bm h_j = \sum_{i=1}^k \lambda_i \bm p_i,
	$$
	where $(*)$ follows from the decomposition property of heterogeneous, isolated simplices (main text, section 2.2). With the same reasoning obtain
	$$
	\sum_{i=1}^k \lambda_i = \sum_{i=1}^{k}\sum_{j\in \Lambda_i}\tilde \lambda_j = \sum_{j=1}^N \tilde \lambda_j = 1,
	$$
	hence $\bm a \in S$ with barycentric coordinates $(\lambda_1, \dots, \lambda_k)$.
	For uniqueness of $S$, let $S' = \text{conv}(\{\bm{p'}_1, \dots, \bm{p'}_k\})$ with $\bm{p'}_i = \sum_{j\in\Lambda_i}\alpha'_j\bm{h}_j$ be another phase split simplex for $\bm a$ with barycentric coordinates $(\lambda'_1, \dots, \lambda'_k)$ (with respect to $S'$). As
	$$
	\sum_{i=1}^k\sum_{j \in \Lambda_i} \lambda'_i\alpha'_j \bm h_j = \bm a = \sum_{i=1}^k\sum_{j \in \Lambda_i} \lambda_i\alpha_j \bm h_j,
	$$
	it must hold that $\lambda'_i\alpha'_j = \lambda_i\alpha_j$ for all $i\in\{1, \dots, k\}$ and $j \in \Lambda_i$ by the uniqueness of the barycentric coordinates of $\bm a$ with respect to $H$. Hence,
	$$
	\lambda_i' = \lambda_i' \underbrace{\sum_{j \in \Lambda_i}\alpha'_j}_{=1} = \sum_{j \in \Lambda_i}\lambda_i'\alpha'_j = \sum_{j \in \Lambda_i}\lambda_i\alpha_j = \lambda_i
	$$
	so $\alpha_j = \alpha'_j$ for all $i$ and $j \in \Lambda_i$, and it follows $S = S'$.
\end{proof}

\section{Results for ternary and quaternary systems}
\label{tern_quat_results}
Here we report results obtained for all ternary and quaternary systems, which were examined in \citep{chen2000, chen2001, yuan2018, yuan2019, yuan2020}. Additionally to MD, we report the computation time (CT), which was needed for our implementation of the CEM to construct the respective phase equilibria. For ternary systems, the classification of the simplices of the convex envelope was executed serially. For systems containing more than three components, the classification step was executed parallelized (the respective settings for parallelization are published alongside the code on GitHub). Table \ref{tern_quant} shows results for several ternary systems containing combinations of $n$-hexane, benzene, sulfolane, toluene, xylene, and $n$-octane. NRTL parameters and experimental data regarding those systems were taken from \citep{chen2000}. As can be seen, the calculated phase splits match the experimental data quite well. Table \ref{quat_quant} shows results for quaternary systems from various sources \citep{chen2000, chen2001, yuan2018, yuan2019, yuan2020}. NRTL and UNIQUAC parameters were used to calculate the occurring phase splits. Compared to ternary systems, $\delta$ was set to a lower value to limit the computational effort. Still, our approach was able to calculate the experimental data with excellent accuracy. Table \ref{quin_quant} shows results for quinary systems from \citep{chen2000, chen2001} using the NRTL model. The CEM was able to calculate the occurring phase splits with high accuracy. Table \ref{six_quant} shows results for systems containing six components, as presented in \citep{yuan2018, yuan2019, yuan2020}. The discretization parameter $\delta$ was set to a lower value to control computational complexity. Still, the CEM was able to calculate the occurring phase splits with an acceptable accuracy.

\begin{table}
	\caption{Results for ternary systems from \citep{chen2000} at atmospheric pressure. The parameter $\delta$ was set to 128 and the NRTL model was used for all systems. $M$ describes the number of feed streams, which were examined for the calculation of the MD.}
	\label{tern_quant}
	\begin{center}
		\begin{tabular}{|p{5.5cm}||c|c|c|c|}
			\hline
			System & $T$ $\backslash$ K & $M$ & CT $\backslash$ s & MD \\
			\hline
			$n$-hexane -- benzene -- sulfolane & 298.15 & 10 & 1.8 & 0.005 \\
			\hline
			$n$-hexane -- toluene -- sulfolane & 298.15 & 10 & 1.7 & 0.004 \\
			\hline
			$n$-hexane -- xylene -- sulfolane & 298.15 & 10 & 1.7 & 0.006 \\
			\hline
			$n$-octane -- benzene -- sulfolane & 298.15 & 10 & 1.7 & 0.005 \\
			\hline
			$n$-octane -- toluene -- sulfolane & 298.15 & 10 & 1.7 & 0.009 \\
			\hline
			$n$-octane -- xylene -- sulfolane & 298.15 & 10 & 1.7 & 0.005 \\
			\hline
		\end{tabular}
	\end{center}
\end{table}

\clearpage
\begin{longtable}{|p{5.5cm}||c|c|c|c|c|}
	\caption{Results for quaternary systems at atmospheric pressure. The parameter $\delta$ was set to 64 for all systems. $M$ describes the number of feed streams, which were examined for the calculation of the MD.} 
	\label{quat_quant} \\
	\hline
	System & $T$ $\backslash$ K & $M$ & $g^E$-model & CT $\backslash$ s & MD \\
	\hline
	\endfirsthead
	\caption*{Table \ref{quat_quant} (continued): Results for quaternary systems at atmospheric pressure. The parameter $\delta$ was set to 64 for all systems. $M$ describes the number of feed streams, which were examined for the calculation of the MD.}
	\endhead
	$n$-hexane -- benzene -- xylene -- sulfolane \citep{chen2000} & 298.15 & 5 & NRTL & 41.6 & 0.003 \\
	\hline
	$n$-hexane -- $n$-octane -- benzene -- sulfolane \citep{chen2000} & 298.15 & 5 & NRTL & 39.2 & 0.005 \\
	\hline
	$n$-octane -- toluene -- xylene -- sulfolane \citep{chen2000} & 298.15 & 5 & NRTL & 41.4 & 0.004 \\
	\hline
	heptane -- benzene -- toluene -- sulfolane \citep{chen2001} & 298.15 & 5 & NRTL & 41.6 & 0.011 \\
	\hline
	heptane -- octane -- $m$-xylene -- sulfolane \citep{chen2001} & 298.15 & 5 & NRTL & 39.7 & 0.007 \\
	\hline
	hexane -- heptane -- toluene -- sulfolane \citep{chen2001} & 298.15 & 5 & NRTL & 38.2 & 0.005 \\
	\hline
	ethanol -- heptanol -- decane -- water \citep{yuan2018} & 298.15 & 21 & UNIQUAC & 49.4 & 0.007 \\
	\hline
	ethanol -- heptanol -- undecane -- water \citep{yuan2018} & 298.15 & 22 & UNIQUAC & 50.4 & 0.008 \\
	\hline
	ethanol -- hexanol -- decane -- water \citep{yuan2018} & 298.15 & 24 & UNIQUAC & 48.6 & 0.005 \\
	\hline
	ethanol -- hexanol -- undecane -- water \citep{yuan2018} & 298.15 & 29 & UNIQUAC & 49.4 & 0.007 \\
	\hline
	ethanol -- nonanol -- dodecane -- water \citep{yuan2019} & 293.15 & 21 & NRTL & 49.6 & 0.016 \\
	\hline
	ethanol -- nonanol -- dodecane -- water \citep{yuan2019} & 298.15 & 13 & NRTL & 50.5 & 0.016 \\
	\hline
	ethanol -- nonanol -- dodecane -- water \citep{yuan2019} & 303.15 & 17 & NRTL & 48.8 & 0.011 \\
	\hline
	ethanol -- nonanol -- tridecane -- water \citep{yuan2019} & 293.15 & 21 & NRTL & 49.8 & 0.009 \\
	\hline
	ethanol -- nonanol -- tridecane -- water \citep{yuan2019} & 298.15 & 18 & NRTL & 49.7 & 0.013 \\
	\hline
	ethanol -- nonanol -- tridecane -- water \citep{yuan2019} & 303.15 & 14 & NRTL & 49.6 & 0.009 \\
	\hline
	ethanol -- octanol -- dodecane -- water \citep{yuan2019} & 293.15 & 22 & NRTL & 46.4 & 0.011 \\
	\hline
	ethanol -- octanol -- dodecane -- water \citep{yuan2019} & 298.15 & 16 & NRTL & 49.6 & 0.013 \\
	\hline
	ethanol -- octanol -- dodecane -- water \citep{yuan2019} & 303.15 & 17 & NRTL & 49.6 & 0.009 \\
	\hline
	ethanol -- octanol -- tridecane -- water \citep{yuan2019} & 293.15 & 21 & NRTL & 48.5 & 0.009 \\
	\hline
	ethanol -- octanol -- tridecane -- water \citep{yuan2019} & 298.15 & 14 & NRTL & 48.7 & 0.013 \\
	\hline
	ethanol -- octanol -- tridecane -- water \citep{yuan2019} & 303.15 & 12 & NRTL & 48.5 & 0.012 \\
	\hline
	ethanol -- decanol -- pentadecane -- water \citep{yuan2020} & 293.15 & 23 & UNIQUAC & 49.7 & 0.006 \\
	\hline
	ethanol -- decanol -- pentadecane -- water \citep{yuan2020} & 298.15 & 22 & UNIQUAC & 49.5 & 0.006 \\
	\hline
	ethanol -- decanol -- pentadecane -- water \citep{yuan2020} & 303.15 & 24 & UNIQUAC & 49.7 & 0.007 \\
	\hline
	ethanol -- decanol -- tetradecane -- water \citep{yuan2020} & 293.15 & 21 & UNIQUAC & 47.8 & 0.007 \\
	\hline
	ethanol -- decanol -- tetradecane -- water \citep{yuan2020} & 298.15 & 24 & UNIQUAC & 49.4 & 0.007 \\
	\hline
	ethanol -- decanol -- tetradecane -- water \citep{yuan2020} & 303.15 & 23 & UNIQUAC & 49.5 & 0.006 \\
	\hline
	ethanol -- undecanol -- pentadecane -- water \citep{yuan2020} & 293.15 & 24 & UNIQUAC & 49.4 & 0.007 \\
	\hline
	ethanol -- undecanol -- pentadecane -- water \citep{yuan2020} & 298.15 & 21 & UNIQUAC & 49.7 & 0.007 \\
	\hline
	ethanol -- undecanol -- pentadecane -- water \citep{yuan2020} & 303.15 & 22 & UNIQUAC & 49.5 & 0.006 \\
	\hline
	ethanol -- undecanol -- tetradecane -- water \citep{yuan2020} & 293.15 & 23 & UNIQUAC & 49.4 & 0.006 \\
	\hline
	ethanol -- undecanol -- tetradecane -- water \citep{yuan2020} & 298.15 & 23 & UNIQUAC & 49.1 & 0.007 \\
	\hline
	ethanol -- undecanol -- tetradecane -- water \citep{yuan2020} & 303.15 & 23 & UNIQUAC & 48.0 & 0.007 \\
	\hline
\end{longtable}

\begin{table}
	\caption{Results for quinary systems at atmospheric pressure. The parameter $\delta$ was set to 32 and the NRTL model was used for all systems. $M$ describes the number of feed streams, which were examined for the calculation of the MD.}
	\label{quin_quant}
	\begin{center}
		\begin{tabular}{|p{5.5cm}||c|c|c|c|}
			\hline
			System & $T$ $\backslash$ K & $M$ & CT $\backslash$ s & MD \\
			\hline
			$n$-hexane -- $n$-octane -- benzene -- toluene -- sulfolane \citep{chen2000} & 298.15 & 4 & 170.4 & 0.010 \\
			\hline
			heptane -- octane -- benzene -- $m$-xylene -- sulfolane \citep{chen2001} & 298.15 & 5 & 161.0 & 0.012 \\
			\hline
			hexane -- heptane -- toluene -- $m$-xylene -- sulfolane \citep{chen2001} & 298.15 & 5 & 165.6 & 0.015 \\
			\hline
		\end{tabular}
	\end{center}
\end{table}

\begin{table}
	\caption{Results for systems containing six components at atmospheric pressure. The parameter $\delta$ was set to 16 for all systems. $M$ describes the number of feed streams, which were examined for the calculation of the MD.}
	\label{six_quant}
	\begin{center}
		\begin{tabular}{|p{5.5cm}||c|c|c|c|c|}
			\hline
			System & $T$ $\backslash$ K & $M$ & $g^E$-model & CT $\backslash$ s & MD \\
			\hline
			ethanol -- hexanol -- heptanol -- decane -- undecane -- water \citep{yuan2018} & 298.15 & 22 & UNIQUAC & 3019.5 & 0.032 \\
			\hline
			ethanol -- octanol -- nonanol -- dodecane -- tridecane -- water \citep{yuan2019} & 293.15 & 20 & NRTL & 3000.0 & 0.036 \\
			\hline
			ethanol -- octanol -- nonanol -- dodecane -- tridecane -- water \citep{yuan2019} & 298.15 & 22 & NRTL & 2960.1 & 0.028 \\
			\hline
			ethanol -- octanol -- nonanol -- dodecane -- tridecane -- water \citep{yuan2019} & 303.15 & 18 & NRTL & 2884.6 & 0.029 \\
			\hline
			ethanol -- decanol -- undecanol -- tetradecane -- pentadecane -- water \citep{yuan2020} & 293.15 & 20 & UNIQUAC & 2245.7 & 0.021 \\
			\hline
			ethanol -- decanol -- undecanol -- tetradecane -- pentadecane -- water \citep{yuan2020} & 298.15 & 24 & UNIQUAC & 2224.7 & 0.020 \\
			\hline
			ethanol -- decanol -- undecanol -- tetradecane -- pentadecane -- water \citep{yuan2020} & 303.15 & 23 & UNIQUAC & 1438.2 & 0.020 \\
			\hline
		\end{tabular}
	\end{center}
\end{table}

\end{document}